\documentclass{emulateapj}

\slugcomment{accepted for ApJ}

\shorttitle{Wide companion near the deuterium burning mass limit
}
\shortauthors{B\'ejar, V. J. S. et al.}

\begin{document}

%% LaTeX will automatically break titles if they run longer than
%% one line. However, you may use \\ to force a line break if
%% you desire.

\title{Discovery of a wide companion near the deuterium burning mass limit 
in the Upper Scorpius association}

\author{V. J. S. B\'ejar\altaffilmark{1}, M. R. Zapatero Osorio\altaffilmark{1}, 
A. P\'erez-Garrido\altaffilmark{2}, C. \'Alvarez\altaffilmark{3}, 
E. L. Mart\'{\i}n\altaffilmark{1,4}, R. Rebolo\altaffilmark{1,5}, 
I. Vill\'o-P\'erez\altaffilmark{2}, A. D\'{\i}az-S\'anchez\altaffilmark{2}}

%% Notice that each of these authors has alternate affiliations, which
%% are identified by the \altaffilmark after each name.  Specify alternate
%% affiliation information with \altaffiltext, with one command per each
%% affiliation.

\altaffiltext{1}{Instituto de Astrof\'{\i}sica de Canarias. 
C/ V\'{\i}a L\'actea s/n, La Laguna, Tenerife, E-38200, Spain. E-mail: vbejar@iac.es}
\altaffiltext{2}{Universidad Polit\'ecnica de Cartagena. 
Campus Muralla del Mar, Cartagena, Murcia, E-30202, Spain.}
\altaffiltext{3}{GTC project. Instituto de Astrof\'{\i}sica de Canarias. 
C/ V\'{\i}a L\'actea s/n, La Laguna, Tenerife, E-38200, Spain.}
\altaffiltext{4}{University of Central Florida. Department of Physics, P.O. 
Box 162385, Orlando, FL 32816-2385, USA.}
\altaffiltext{5}{Consejo Superior de Investigaciones Cient\'{\i}ficas, Spain.}

%% Mark off your abstract in the ``abstract'' environment. In the manuscript
%% style, abstract will output a Received/Accepted line after the
%% title and affiliation information. No date will appear since the author
%% does not have this information. The dates will be filled in by the
%% editorial office after submission.

\begin{abstract}
We present the discovery of a companion near the deuterium burning mass limit located at a very wide distance, at an 
angular separation of 4.6$\pm$0.1\arcsec\,(projected distance of $\sim$\,670 AU) 
%and position angle of 177$\pm$1$\degr$ 
from UScoCTIO\,108, a brown dwarf of the very young 
Upper Scorpius association. Optical and near-infrared photometry and spectroscopy 
confirm the cool nature of both objects, with spectral types of M7 and M9.5, respectively, 
and that they are bona fide members of the association, showing
low gravity and features of youth. Their masses, estimated from the comparison of 
their bolometric luminosities and theoretical models 
for the age range of the association, are $60\pm 20$ and $14^{+2}_{-8}$\,M$_{Jup}$, respectively. 
The existence of this object around a brown dwarf at this wide orbit suggests that the companion is unlikely to have  
formed in a disk based on current planet formation models. Because this system is rather weakly bound, they did not probably form 
through dynamical ejection of stellar embryos.

\end{abstract}

%% Keywords should appear after the \end{abstract} command. The uncommented
%% example has been keyed in ApJ style. See the instructions to authors
%% for the journal to which you are submitting your paper to determine
%% what keyword punctuation is appropriate.


%% From the front matter, we move on to the body of the paper.
%% In the first two sections, notice the use of the natbib \citep
%% and \citet commands to identify citations.  The citations are
%% tied to the reference list via symbolic KEYs. The KEY corresponds
%% to the KEY in the \bibitem in the reference list below. We have
%% chosen the first three characters of the first author's name plus
%% the last two numeral of the year of publication as our KEY for
%% each reference.

%% Authors who wish to have the most important objects in their paper
%% linked in the electronic edition to a data center may do so by tagging
%% their objects with \objectname{} or \object{}.  Each macro takes the
%% object name as its required argument. The optional, square-bracket 
%% argument should be used in cases where the data center identification
%% differs from what is to be printed in the paper.  The text appearing 
%% in curly braces is what will appear in print in the published paper. 
%% If the object name is recognized by the data centers, it will be linked
%% in the electronic edition to the object data available at the data centers  
%%
%% Note that for sources with brackets in their names, e.g. [WEG2004] 14h-090,
%% the brackets must be escaped with backslashes when used in the first
%% square-bracket argument, for instance, \object[\[WEG2004\] 14h-090]{90}).
%%  Otherwise, LaTeX will issue an error. 

\section{Introduction}

To date, about 250 extrasolar planets have been identified using the indirect techniques of radial velocity, 
photometric transit and 
microlensing \citep{may95,kon03,uda05}. 
The direct detection of a planet's light allows a  
detailed study of its physical properties and it is crucial to improve our undestanding of these objects. 
At present, one extrasolar planetary-mass companion 
has been directly imaged around the brown dwarf \object{2MASS\,J12073347-3932540} in the very young TW Hydra association 
\citep{cha04}.
Several substellar companions with masses close but likely above the deuterium burning mass limit ($\sim$ 13 $M_{Jup}$\footnote{ 1 M$_{\odot}$ = 1047.7 M$_{Jup}$ }\,, \citealt{sau96}) have also been imaged around 
stellar and substellar primaries \citep{neu05,cha05,ito05,luh04,luh06,jay06,all06,clo07}. 
All these systems are young ($<$ 50\,Myr) and have projected separations lower than 260\,AU. 
%Only two of them, 2MASSJ1207-3932 and Oph 1622-2405, have a brown dwarf primary and their projected separations are 55 and 243\,AU, respectively.
There are also two unconfirmed planetary candidates imaged around brown dwarfs in the $\sigma$ and $\lambda$ Orionis 
clusters \citep{cab06,bar07}.
%Very recently, an amount of very low mass binaries at wide separations ($>$ 100\,AU) have been
%also discovered in very young cluster and associations and in the 
%field (see Caballero \cite{cab07} and references therein). 
Here, we present the discovery of a companion near the deuterium burning mass limit located at the much wider projected separation 
of 670\,AU from the brown dwarf \object{UScoCTIO\,108} \citep*{ard00}, which belongs to the Upper Scorpius (USco) association.  
This is one of the youngest and closest OB associations to the Sun, located at an 
average distance of 145$\pm$2\,pc and with an estimated age of 5--6 Myr, \citep{dez99,pre99}.

%\object[]{UScoCTIO108}!

\section{Observations}

We have searched for very red and faint companions ($J$$>16$ and $J-K_{\rm s}$$>1$)
 around 500 previously known members and candidates of the USco association 
(\citealt{ard00}; \citealt*{pre01}; \citealt{pre02}; \citealt*{mar04}; \citealt{lod07}) using the 2MASS and DENIS catalogues database and the 
United Kingdom Schmidt Telescope (UKST) plates. 
Among other candidates, we identified a red source (\object{2MASS\,J16055409-1818488}) 
in the 2MASS catalogue around UScoCTIO\,108 (\object{2MASS\,J16055407-1818443}). 
Follow-up observations in the $I$-band showed that the new object has $I-J=3.38\pm0.09$ 
and is located at an angular distance of 4.6$\pm$0.1\arcsec\, and a position angle of 177$\pm$1$\degr$. 
Additional optical and near-infrared imaging and low resolution spectroscopy were carried out using 
different instrumentation. The detailed observing log is provided in Table~\ref{tab1}.  
%Follow-up observations in the $I$-band obtained on 2007 July 5, using the CCD camera
%(2k$\times$2k, 0.305$"$\,pix$^{-1}$) mounted on the IAC-80 at the Teide Observatory, showed that the new  
%object has a very red $I-J$ color and that it is located at an angular distance of 4.7\,arcsec. 
%On 2007 July 15, we also obtained $I$ and $Z$-band images using the AUX camera (1k$\times$1k,
%0.108$"$\,pix$^{-1}$) mounted on the Cassegrain focus of the 4.2\,m William Herschel Telescope (WHT) at the Observatorio 
%Roque de los Muchachos (ORM) with a total integration time of 600\,s in each filter. 
%During the same night, we performed low resolution optical spectroscopy of both objects, using 
%the red arm of the ISIS instrument (WHT) and the R158R grating, providing a nominal dispersion of 1.8\,\AA\,pix$^{-1}$.  
%The total integration time was 7200\,s and the resolution of the spectrum 6\,\AA. 
%We also acquired near-infrared images in the $J$, $H$, and $K'$ bands, with a total exposure time of 300\,s in each filter, 
%using the NICS camera (1k$\times$1k, 0.25"\,pix$^{-1}$) mounted on the Telescopio Nationale Galileo (TNG) at the ORM on the 2007 July 16. 
%We also obtained $J$-band spectroscopy (1.14--1.35~$\mu$\,m), using the cross-dispersion spectrograph  NIRCSPEC 
%and nominal dispersion of 2.8\,\AA\,pix$^{-1}$ at the 10\,m KeckII Telescope on 2007 July 24. Total integration time was 600\,s and the 
%resolution of the spectrum 9\,\AA. 
Weather conditions were photometric at all the observatories and average seeing was in the range 0.7--1.0\arcsec.

\begin{figure}
\epsscale{.80}
\plotone{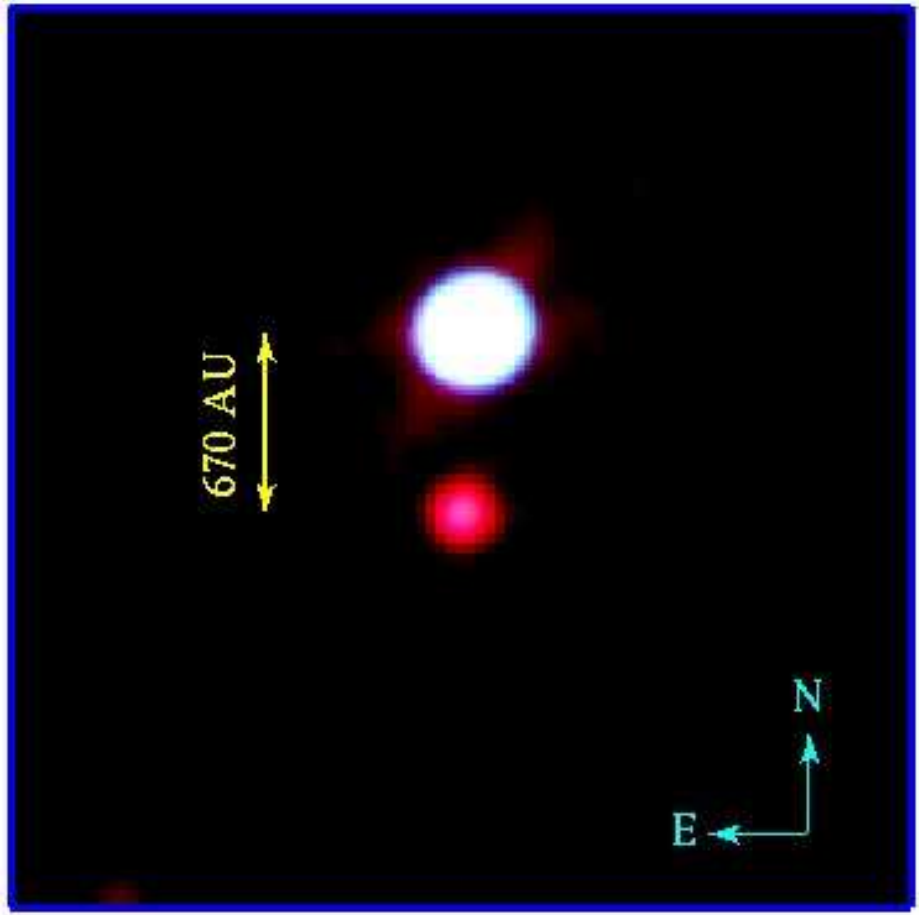}
\caption{Composite colour $IZK$ image (19.5\arcsec$\times$19.5\arcsec) of UScoCTIO 108A and B ($I$ = blue, $Z$ = green, $K'$ = red). 
$IZ$ images are from AUX intrument on the WHT and the $K'$ image from NICS on the TNG. All images were convolved with a
gaussian to a spatial resolution of 1.1\arcsec.\label{fig1}}
\end{figure}

We reduced the data with standard techniques, using 
routines within the IRAF environment, including bias and flat-field correction in the
 optical, and sky subtraction and flat-field correction in the near infrared. 
Finally, we aligned and combined individual images to obtain the final one. 
A composite colour image using the $IZK'$ bands of UScoCTIO\,108A and B is shown in 
Figure~\ref{fig1}. 
We have performed aperture and PSF photometry of the resulting images using the DAOPHOT 
package. Optical and near-infrared images have been calibrated using bright sources in common with 
the DENIS and 2MASS catalogues. The photometry of both the primary and
the secondary are indicated in Table~\ref{tab2}. 
The spectra were extracted using the APALL routine,  wavelength calibrated, and corrected for 
the instrumental response with observations of the spectrophotometric standard stars \object{Wolf1346} and \object{HZ44}. 
Near-infrared spectra have been corrected for telluric lines by dividing them by the A3-type star 
\object{HD\,142613} and multiplying by a blackbody of the corresponding effective temperature ($T_{\rm eff}$) of 8500\,K.

\section{Physical properties and membership of USco}

\begin{figure}
\epsscale{.80}
\plotone{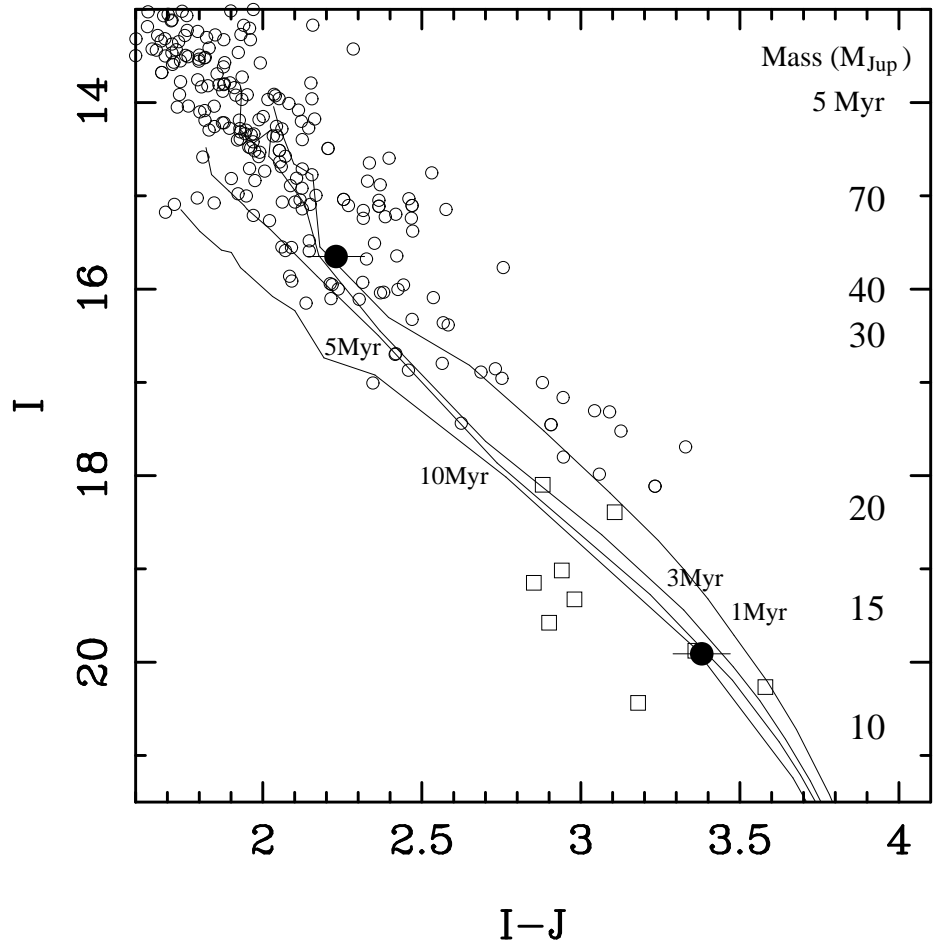}
\caption{$I, I-J$ color--magnitude diagram of previously known members of the USco association (open circles) 
and isolated planetary-mass objects in the $\sigma$ Orionis cluster (open squares) shifted to the distance of the former. 
UScoCTIO 108A and B are represented by solid circles with their photometric error bars. 
The isochrones from the Lyon group models \citep{bar03} and estimated masses for the 5\,Myr age 
(in Jupiter mass units) are also indicated.\label{fig2}}
\end{figure}

Using the photometry from Table~\ref{tab2}, we have determined that UScoCTIO\,108A and B 
belong to the photometric sequence of the USco association. In Figure~\ref{fig2}, we represent an
$I$, $I-J$ color--magnitude diagram, 
where both objects are indicated by filled circles. The primary follows the sequence of previously known members 
\citep{ard00,pre01,pre02,mar04,lod07} and the secondary smoothly extrapolates it towards fainter magnitudes and 
redder colours. We note that UScoCTIO\,108B lies on the location of isolated planetary-mass objects 
in the $\sigma$ Orionis cluster \citep{zap00} when they are shifted to the distance of the USco association (see
Figure~\ref{fig2}). 
Using the WHT and IAC80 $I$-band images and the astrometry provided by 2MASS, we have measured  the proper motion 
of UScoCTIO\,108A to be ($\mu_\alpha$\,cos$\delta$, $\mu_\delta$ = $-8\pm$14, $-17\pm$13)\,mas\,yr$^{-1}$ and UScoCTIO\,108B to be 
($\mu_\alpha$\,cos$\delta$, $\mu_\delta$ = $-6\pm$40, $-20\pm$40)\,mas\,yr$^{-1}$. Both measurements are consistent with the 
proper motion of the USco association ($\mu_\alpha$\,cos$\delta$, $\mu_\delta$ = $-11$, $-25$\,mas\,yr$^{-1}$, 
\citealt{dez99}), 
but the large error bars of the secondary prevent us from  reaching any firm conclusion. 

\begin{figure}
\epsscale{1.6}
\plotone{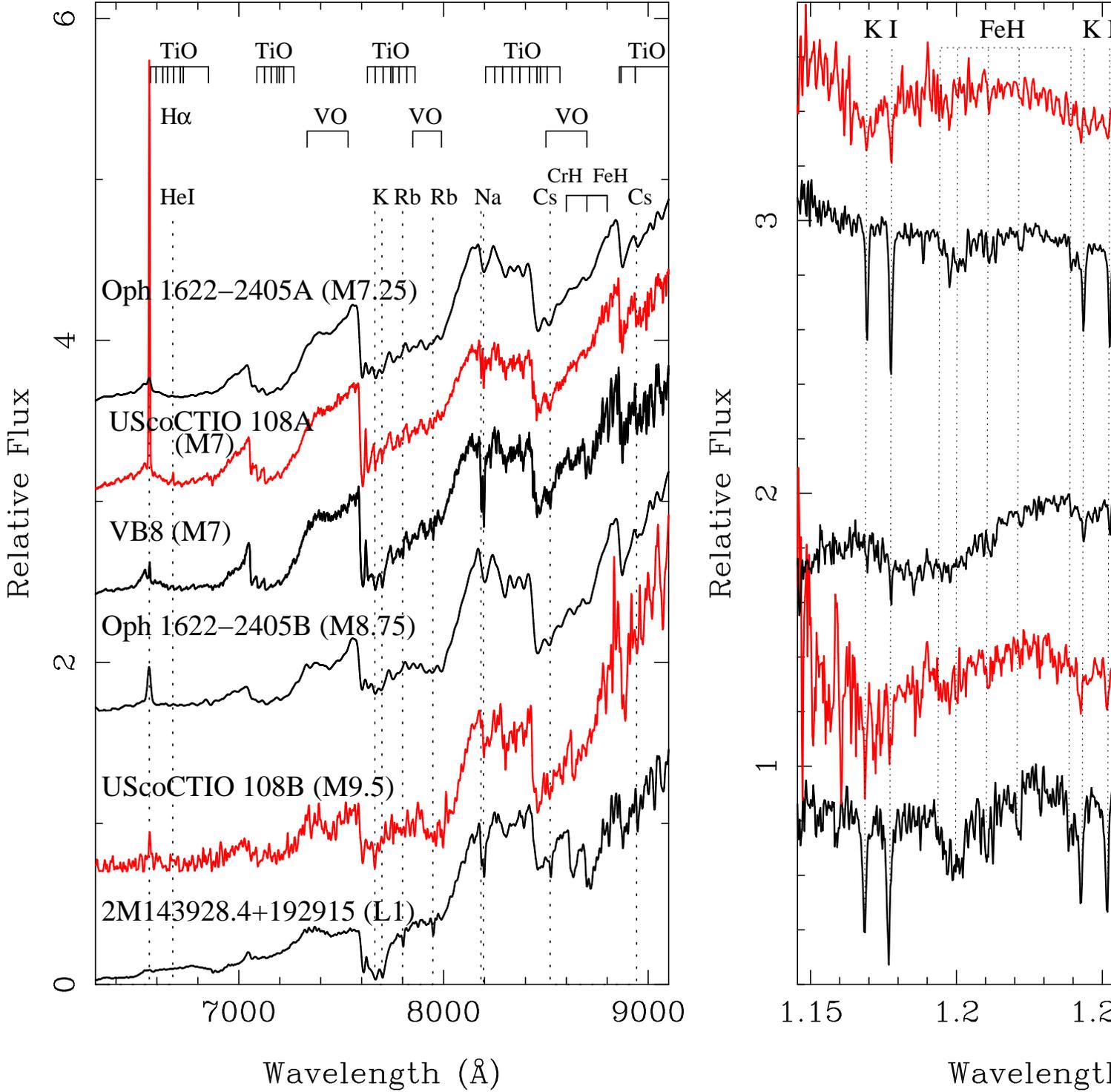}
\caption{Optical (left panel) and $J$-band (right panel) spectra of UScoCTIO 108A and B 
(in red in the electronic version), young objects (data from \citealt{luh07} and \citealt{mcg04}) 
and field dwarfs (from this paper, \citealt{kir99}, and \citealt{mcl03}) of a similar spectral type. Their names, spectral type, 
and main spectroscopic features are indicated. All the spectra have been normalized to unity at 8175\AA\, 
and 1.30~$\mu$m.\label{fig3}}
\end{figure}

We have determined the spectral classification of UScoCTIO\,108A and B by comparison with 
standard objects of well-known spectral type and using PC3 and PC4 indexes for the optical spectra \citep*{mar96} 
and the water index at 1.2~$\mu$m for the infrared ones \citep{geb02}. 
In Figure~\ref{fig3}, we present our spectra and data of other young (\object{Oph\,1622-2405AB}, \object{KPNO\,Tau4}) 
and field dwarfs (\object{VB8}, \object{2MASS\,J1439284+192915}, \object{2MASS\,J1506544+132106}) taken from the 
literature (see references in the caption of the figure). 
%optical spectra (left panel) of UScoCTIO\,108A and B 
%in comparison with the M7 and L1 field dwarfs \object{VB8} and \object{2MASS\,J1439284+192915} (data 
%from \citealt{kir99}), and the young M7.25/M8.75 binary \object{Oph\,1622-2405AB} (from \citealt{luh07}) 
%and their near-infrared spectra (left panel) in comparison with \object{VB8},  
%the L3 field dwarf \object{2MASS\,J1506544+132106} (from \citealt{mcl03}) and the young M9.5 \object{KPNO\,Tau4} 
%(from \citealt{mcg04}). 
We have derived a spectral type for the primary of M7 with an error of half a subclass in both 
the optical and near-infrared spectra, although the bluer part of the optical spectrum seems to be 
of a hotter object. 
We estimate that the secondary is an M9.5 by comparison of the optical and $J$-band data with the 
young sources Oph\,1622-2405B and KPNO\,Tau4, an L1 according to the pseudocontinuum 
index PC3 ([823--827]/[754--758]\,nm), and 
its near-infrared spectrum is similar to the L3 field dwarf 2MASS\,J1506544+132106.  
The slightly different typings may be due to the effects of a low-pressure, cool atmosphere on the various optical and near-infrared spectroscopic features. 
We finally adopt a classification of M9.5 for UScoCTIO\,108B. 
The spectral types of the primary and secundary correspond to a 
$T_{\rm eff}$ of 2700$\pm$100\,K and 2350$^{+100}_{-400}$\,K, respectively, adopting the temperature 
scale for high gravity field dwarfs \citep{dah02,gol04}. 
The $T_{\rm eff}$ of UScoCTIO\,108A is consistent with the $T_{\rm eff}$ calculated for both components of 
the low-gravity M6.5 eclipsing binary \object{2MASS\,J05352184-0546085} (2900 and 2800\,K). 
We have computed these values from the total luminosity (estimated from the $K$-band magnitude, the bolometric 
correction from \citealt{gol04}, and a distance of 480\,pc), the radii, and $T_{\rm eff}$ ratio given by \cite{sta06}. 

Optical spectroscopy of UScoCTIO\,108A shows spectral features characteristic of youth, such as a very strong H$_{\alpha}$ 
(equivalent width, EW $=-90\pm2$\,\AA) and He\,{\sc i} emission lines (EW[5876\AA]$=-10\pm2$\,\AA, EW[6678\AA]$=-1.5\pm0.5$\,\AA), 
which indicate that the primary is still in the process of accreting  
from a disk. In addition, alkaline lines such as Na\,{\sc i} and K\,{\sc i} are weaker than their field dwarf 
counterparts, which is 
characteristic of still contracting low-gravity objects. The Li\,{\sc i} line is also detected in absorption (EW $=0.45\pm 0.1$\,\AA), 
but it is slightly less intense than expected for its spectral type and youth. This could be caused by the higher continuum 
in this region, i.e., veiling, probably due to the accretion of material from the disk. By dividing its spectrum by 
that of other non-accreting M7 dwarfs, such as \object{SOri\,27} \citep{zap02}, \object{SOri\,40} \citep{bej99} and \object{VB8} (this paper), we estimate a veiling factor ($r=F_{\rm excess}/F_{\rm photo}$) of 
0.4--0.7, which gives a corrected EW(Li\,{\sc i}) = 0.6--0.8\,\AA, consistent with a total preservation of this element.  
Optical spectroscopy of UscoCTIO\,108B also shows H$_{\alpha}$ in emission, but this is less intense than in the primary
(EW $=-15\pm10$\,\AA). The presence of this line is rare (less than 20\%) in the spectra of field L dwarfs \citep{sch07}, and this could be a signature of youth. 
The Na\,{\sc i} and K\,{\sc i} lines are weaker and the TiO and VO molecular bands are more intense than expected for objects of the same 
spectral type in the field (see Figure~\ref{fig3}), which are also indicative of youth. 
%This optical spectrum is very similar to that recently presented by \cite{kir06} of a very young L dwarf. 
Low-gravity features are even clearer in the $J$-band spectra, where 
both the primary and secondary show weaker K lines than the late-type field dwarfs. 
The hydrides (FeH and CrH) also appear weaker at optical and near-infrared wavelengths in the USco objects than in the field dwarfs. 
This is likely related to an intense TiO absorption characteristic of low gravity, cool atmospheres \citep{mar96}.  
%We have identified another spectroscopic feature in the $J$-band spectra that could be indicative of low pressures (hence, 
%low gravities): the USco objects show stronger, moderately  broad absorption at around 1.27 $\mu$m, which includes the 
%Na\,{\sc i} line at 1.26781 $\mu$m. Neither the broad absorption nor the atomic line are apparent in the low-resolution data 
%of the high-gravity dwarfs.
In summary, from optical and near-infrared spectroscopy, we may conclude that UScoCTIO\,108A and B have  spectral features 
of a very young age, which support their membership of the USco association. 

\begin{figure}
\epsscale{.80}
\plotone{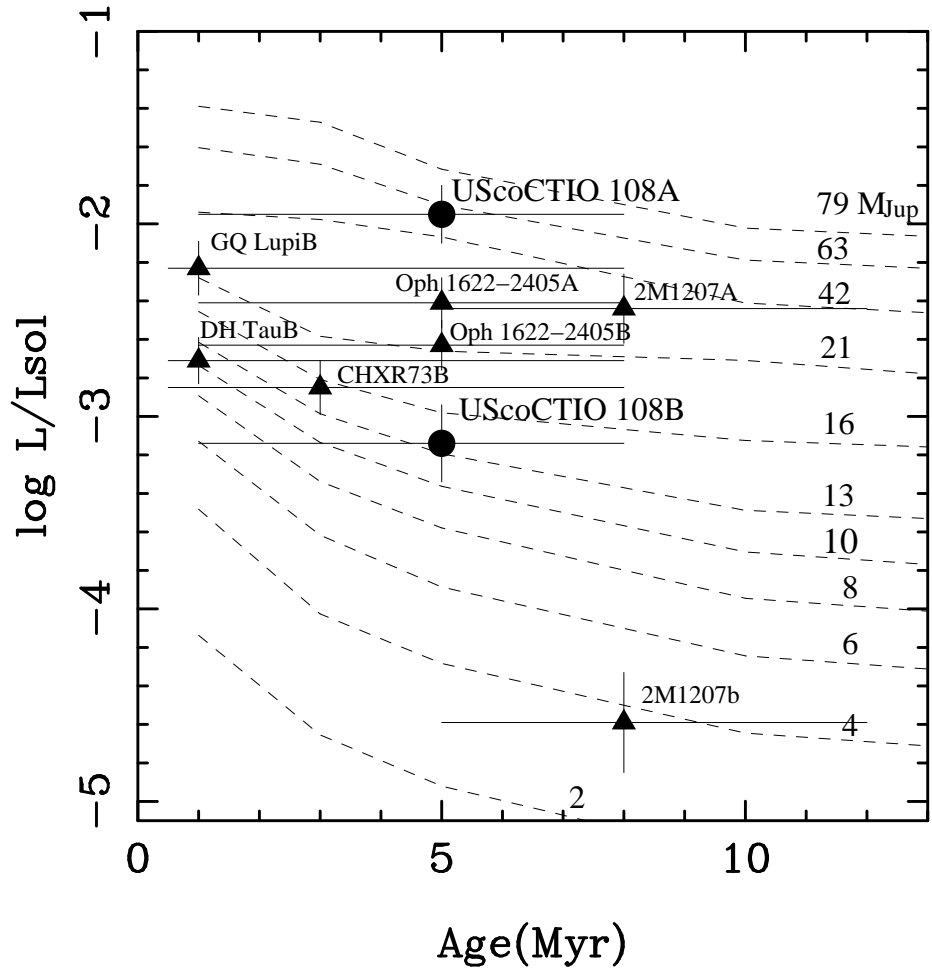}
\caption{Luminosity, age diagram. UScoCTIO108\,A and B, are represented by solid circles, and other very low-mass 
substellar companions as triangles (data from \citealt{luh06,luh07,cha04}). Evolutionary tracks (dashed lines) 
from the Lyon group models \citep{bar03} are also shown. Masses to the right are indicated in Jupiter mass
units.\label{fig4}}
\end{figure}

We have derived the luminosity of both objects from their $IJK'$-band magnitudes, the bolometric correction 
from \cite{dah02}, \cite{gol04} and the USco distance modulus $m-M=5.81\pm0.3$ \citep{dez99}. 
We have not applied any reddening correction to apparent magnitudes since the extinction in  
the USco association is found to be quite small ($A_{V}<2$, \citealt{pre99}). 
We have obtained a luminosity of $\log L/L_{\odot}=-1.95^{+0.17}_{-0.15}$ for UScoCTIO\,108A and 
$\log L/L_{\odot}=-3.14\pm 0.20$  for UScoCTIO\,108B. We can estimate the mass of the objects by comparison of the derived 
luminosity with predictions from theoretical models \citep{bar03,bur97}. Figure~\ref{fig4} shows the luminosity
of both objects and other very low-mass substellar companions  
in comparison with evolutionary models from \cite{bar03}. 
Isochrone fitting to the more massive stars sequence suggests an age of 5--6\,Myr for USco \citep{pre99}. From this estimated
age, we obtain a mass of $60\pm 10\,M_{\rm Jup}$ for the primary, i.e., within the brown dwarf domain, and 
a mass of $14\pm 2\,M_{\rm Jup}$ for the secondary, i.e., at the deuterium burning mass limit. 
The existence of Li in very low-mass stars provide an alternative way to restrict the age of the association, because
this element is destroyed very fast in their fully convective interiors. 
The comparison of the Li abundance in early M-type members \citep{pre01} with theoretical spectral synthesis \citep{zap02} indicates 
that most of them preserve their initial Li content. According to evolutionary models,  
%and following the same argumentation as given in \cite{zap02} for the $\sigma$ Orionis cluster, 
this indicates that the age of the association is lower than 8\,Myr and most likely in
the interval 2--4\,Myr (see \citealt{zap02}). In fact, the great similarity between the photometric sequences of USco and $\sigma$ Orionis, 
which has a likely age of 3 Myr, when both star associations have been moved to the same distance, and 
UScoCTIO\,108A being still in a strong accretion phase, also argue in favor of a younger age for the system. 
%For this age, we derive a mass of $45\pm 10\,M_{\,Jup}$ for the primary, and a mass of $11\pm 2\,M_{\,Jup}$ for the
%secondary, i.e., within the planetary-mass domain. 
Adopting the wider range of ages of 1--8\,Myr, we estimate a conservative wider mass range for both components: 
$60\pm 20$\,$M_{\rm Jup}$ for the primary and $14^{+2}_{-8}$\,$M_{\rm Jup}$ for the secondary. 
%From Figure~\ref{fig4}, we can see that UScoCTIO\,108B has the lowest luminosity and probably the lowest mass, 
%after \object{2MASS\,J12073347-3932540b}. 
%The comparison of the $T_{\rm eff}$ and the observed magnitudes with the values predicted by the 
%evolutionary models give similar, but slightly lower, masses ($12-13\,M_{\rm Jup}$ for the secondary at 5\,Myr)}.

%Other methods to determine masses 
%{by comparison of the $T_{\rm eff}$ or magnitudes with those derived from models give similar, but slightly lower, 
%masses ($12-13\,M_{\rm Jup}$ for the secondary at 5\,Myr)}. 

\section{Evidence of binarity}

Once we have demonstrated that UScoCTIO\,108A and B are members of the USco association 
and we have estimated their masses, one 
question  still remains open, which is whether the binary is physically bound or just a chance projection effect. 
To check this, we have estimated the probability of finding a planetary-mass member ($J>16$) in our search 
within a radious of 10\arcsec around 500 members and candidates of the association. This exploration is limited  
by sensitivity of the 2MASS Point Source Catalogue \citep{cut03}, which is $J\sim17-17.5$. 
We have derived  the density of such objects with a $J$-band magnitude
in the range of 16--17.5 to be $\rho\sim1.1$\,deg$^{-2}$ from a survey \citep{lod07} that is much deeper than  2MASS. Assuming a Poissonian 
distribution for the number of additional members in a given area, 
we can estimate this probability to be 1-$P$, and $P=P(x=0)=\exp(-np)$, where $P$ is the probability of finding no additional member, 
$n$ is the number of events (500), and $p$ is the expected number of objects in a 10\arcsec radious ($p=\rho*{\rm area}=2.67*10^{-5}$), with the result that there is a probability 
of  about 1.3\% that UScoCTIO\,108B is another member of the association located by chance in the direction of UScoCTIO\,108A. 
If we consider only the probability of finding another member at the distance of UScoCTIO\,108B (4.6\arcsec), 
this probability turns out to be lower by a factor of 4.5.
%From near-infrared spectra, we have estimated that the relative velocity between both components is of 11$\pm$15 Km/s, consistent 
%with 0. Given the relative large velocity dispersion in the Upper Scorpius association of 59 Km/s ({\it 24}), this also reinforce 
%our claim that UScoCTIO 108A and UScoCTIO 108B are a real binary system physically bound. 

The projected separation of both 
components, 670\,AU for the average distance to the association, is also not very common in very low-mass stars and 
star/brown dwarf systems, but there are some known cases at this separation and even at larger ones (see Fig. 15 from \citealt{clo07}). 
The escape velocity from the primary at this distance is  only 0.4 km\,s$^{-1}$ 
and the gravitational bound energy of the system is 1.86$\times$10$^{33}$\,J. 
%lower than the kinetic energy of the stars in the
%association. This seems to indicate that the binary will become unbound in the near future. 
Although other substellar pairs with similar mass ratio are known, the UScoCTIO\,108A and B system is the 
widest identified so far and possibly has a lowest gravitational bound energy than any other known low-mass
binary (see Fig. 16 from \citealt{clo07}). 
Following the analytical solutions given in \cite{wei87} and \cite{bin87}, we  estimate that the timescale of disruption of the system 
in an environment with the typical density of USco ($\sim$ 0.3 object\,pc$^{-3}$) is  a few hundred  
million years, which is a longer timescale than that expected for the dissipation of USco. 
%For a similar density to that of the 
%solar vicinity ($\sim$ 0.07 object\,pc$^{-3}$), we estimate that UScoCTIO\,108A and B will not be disrupted until several Gyr,  
%so there could be similar binary systems (but much cooler and fainter) at the age of the Sun. 
%The fact that
%UScoCTIO\,108A and B are gravitationally bound at the age of the association ($\sim$5\,Myr) 
%suggests, using the same equations as above, that this system was formed in a low density environment ($<$20 object\,pc$^{-3}$) and
%has not suffered strong interactions with nearby, more massive cluster stars. 

\section{Conclusions and final remarks}

In conclusion, we have found a $14^{+2}_{-8}$\,$M_{\rm Jup}$ companion to the 
$60\pm 20$\,$M_{\rm Jup}$ brown dwarf UScoCTIO\,108 at an angular separation of 4.6$\pm$0.1\arcsec\,
(projected distance of $\sim$\,670 AU) in the very young USco association. 
It seems very difficult to explain the in situ formation of 
this object in a disk by core accretion \citep{pol96} or disk instability \citep{bos97}. Given the 
typical size and density profile of stellar disks, there does not seem  to be enough mass at such a wide separation to form 
a companion of this mass, unless it has  originated at a lower distance and migrated to its present location. 
This very low-mass substellar system has a low binding energy, implying that it is unlikely to have 
been ejected from a higher mass unstable multiple system \citep{rei01}. 
A more likely scenario is that the system was originated from the disruption of a more massive core \citep{bod98} in a way similar  
to other binary stars are supposed to be formed. If the formation of these wide and very low mass systems in the denser 
central part of clusters is relatively frequent, this could explain the existence of isolated 
planetary-mass objects as planetary-mass companions that became unbound from their primaries.
%In addition, the formation of substellar systems at such wide separations 
%seems very difficult to  explain as stellar embryos dynamically ejected from multiple systems 
%at early stages \citep{rei01}, as this process is predicted to produce a lack of very wide binaries.     

\notetoeditor{Would it be possible to include more acknowledgments in the electronic version?}

\acknowledgments

We thank J. Licandro, N. Pinilla-Alonso, J. de Le\'on, M. Montgomery, R. Deshpande, and R. Tata 
for their help in the acquisition of some data, G. Bihain, and 
the referee K. Luhman for his comments and providing his data of Oph\,1622-2405AB. 
We are indebted to T. Mahoney for revising the English  of this manuscript. 

{\it Facilities:} \facility{IAC80 (CCD)}, \facility{WHT (AUX, ISIS)}, \facility{Keck (NIRSPEC)}, 
\facility{TNG (NICS)}.

%% Appendix material should be preceded with a single \appendix command.
%% There should be a \section command for each appendix. Mark appendix
%% subsections with the same markup you use in the main body of the paper.

%% Each Appendix (indicated with \section) will be lettered A, B, C, etc.
%% The equation counter will reset when it encounters the \appendix
%% command and will number appendix equations (A1), (A2), etc.

%\appendix

%\section{Acknowledgments}

\clearpage

%% Use the figure environment and \plotone or \plottwo to include
%% figures and captions in your electronic submission.
%% To embed the sample graphics in
%% the file, uncomment the \plotone, \plottwo, and
%% \includegraphics commands
%%
%% If you need a layout that cannot be achieved with \plotone or
%% \plottwo, you can invoke the graphicx package directly with the
%% \includegraphics command or use \plotfiddle. For more information,
%% please see the tutorial on "Using Electronic Art with AASTeX" in the
%% documentation section at the AASTeX Web site,
%% http://www.journals.uchicago.edu/AAS/AASTeX.
%%
%% The examples below also include sample markup for submission of
%% supplemental electronic materials. As always, be sure to check
%% the instructions to authors for the journal you are submitting to
%% for specific submissions guidelines as they vary from
%% journal to journal.

%% This example uses \plotone to include an EPS file scaled to
%% 80% of its natural size with \epsscale. Its caption
%% has been written to indicate that additional figure parts will be
%% available in the electronic journal.

%\clearpage

\clearpage

%% Here we use \plottwo to present two versions of the same figure,
%% one in black and white for print the other in RGB color
%% for online presentation. Note that the caption indicates
%% that a color version of the figure will be available online.
%%

\clearpage

\clearpage

%% This figure uses \includegraphics to scale and rotate the still frame
%% for an mpeg animation.

%\begin{figure}
%\includegraphics[angle=90,scale=.50]{f3.eps}
%\caption{Animation still frame taken from \citet{kim03}.
%This figure is also available as an mpeg
%animation in the electronic edition of the
%{\it Astrophysical Journal}.}
%\end{figure}

%% If you are not including electonic art with your submission, you may
%% mark up your captions using the \figcaption command. See the
%% User Guide for details.
%%
%% No more than seven \figcaption commands are allowed per page,
%% so if you have more than seven captions, insert a \clearpage
%% after every seventh one.

%% Tables should be submitted one per page, so put a \clearpage before
%% each one.

%% Two options are available to the author for producing tables:  the
%% deluxetable environment provided by the AASTeX package or the LaTeX
%% table environment.  Use of deluxetable is preferred.
%%

%% Three table samples follow, two marked up in the deluxetable environment,
%% one marked up as a LaTeX table.

%% In this first example, note that the \tabletypesize{}
%% command has been used to reduce the font size of the table.
%% We also use the \rotate command to rotate the table to
%% landscape orientation since it is very wide even at the
%% reduced font size.
%%
%% Note also that the \label command needs to be placed
%% inside the \tablecaption.

%% This table also includes a table comment indicating that the full
%% version will be available in machine-readable format in the electronic
%% edition.
\clearpage

\begin{deluxetable}{lcccccccc}
\tabletypesize{\scriptsize}
\tablecaption{Observing log\label{tab1}}
\tablewidth{0pt}
\tablehead{
\colhead{Telescope} & \colhead{Instrument} & \colhead{Mode} & \colhead{Plate Scale}& \colhead{Wavelength range} & \colhead{Dispersion} & \colhead{Resolution} 
& \colhead{Obs. date} & \colhead{Exp. time}\\
                    &                      &                & \colhead{(\arcsec\,pix$^{-1}$)}& \colhead{($\mu$m)}      & \colhead{(\AA\,pix$^{-1}$)} & \colhead{(\AA)} 
&  & \colhead{(s)} 
}
\startdata
IAC80  & CCD (2k$\times$2k)   & Imaging      & 0.305   & $I$	    & \nodata & \nodata &  2007 July 5   &  3600  \\
WHT    & AUX (1k$\times$1k)   & Imaging      & 0.108   & $IZ$	    & \nodata & \nodata &  2007 July 15  &   600  \\ 
WHT    & ISIS (R158R grating) & Spectroscopy & 0.22    & 0.55--0.95 & 1.8     & 6	&  2007 July 15  &  7200  \\ 
TNG    & NICS (1k$\times$1k)  & Imaging      & 0.25    & $JHK'$     & \nodata & \nodata &  2007 July 16  &   300  \\ 
KeckII & NIRSPEC              & Spectroscopy & 0.19    & 1.14--1.35 & 2.8     & 9	&  2007 July 24  &   600  \\ 

\enddata
%% Text for table notes should follow after the \enddata but before
%% the \end{deluxetable}. Make sure there is at least one \tablenotemark
%% in the table for each \tablenotetext.
%%\tablecomments{}
%%\tablenotetext{a}{Sample footnote for table~\ref{tbl-1} that was generated
%%with the deluxetable environment}
\end{deluxetable}

\clearpage

\begin{deluxetable}{lccccccccc}
\tabletypesize{\scriptsize}
\tablecaption{Photometric and spectroscopic data and physical parameters\label{tab2}}
\tablewidth{0pt}
\tablehead{
\colhead{ID} & \colhead{$I$} & \colhead{$I-Z$} & \colhead{$I-J$} & \colhead{$J-H$} & \colhead{$J-K'$} &
 \colhead{Sp. Type} & \colhead{Luminosity} & \colhead{$T_{eff}$} & \colhead{Mass} \\
\colhead{} & \colhead{} & \colhead{} & \colhead{} & \colhead{} & \colhead{} &
 \colhead{} & \colhead{log $L/L_{\odot}$} & \colhead{(K)} & \colhead{(M$_{Jup}$)}
}
\startdata
UScoCTIO\,108A & 15.65$\pm$0.08 & 1.00$\pm$0.1  & 2.23$\pm$0.09 & 0.58$\pm$0.04 & 0.91$\pm$0.04 &    M7     & -1.95$^{+0.17}_{-0.15}$ & 2700$\pm$100          & 60$\pm$20 \\
UScoCTIO\,108B & 19.91$\pm$0.08 & 1.30$\pm$0.1  & 3.38$\pm$0.09 & 0.78$\pm$0.08 & 1.42$\pm$0.11 &   M9.5    & -3.14$\pm$0.20          & 2350$^{+100}_{-400}$  & 14$^{+2}_{-8}$\\
\enddata
%% Text for table notes should follow after the \enddata but before
%% the \end{deluxetable}. Make sure there is at least one \tablenotemark
%% in the table for each \tablenotetext.
%%\tablecomments{}
%%\tablenotetext{a}{Sample footnote for table~\ref{tbl-1} that was generated
%%with the deluxetable environment}
\end{deluxetable}

\end{document}